\documentclass[journal]{vgtc}                     

\usepackage{times}

\usepackage{tabu}                      
\usepackage{booktabs}                  
\usepackage{lipsum}                    
\usepackage{mwe}                      
\usepackage{mathptmx}                  
\usepackage{multirow}
\usepackage{tabularx}
\usepackage{makecell}                      
\usepackage{setspace}             
\usepackage{multirow}
\usepackage{tabu}                                                        
\usepackage{amsmath} 
\usepackage{xcolor}
\usepackage{multirow}
\usepackage{enumitem}
\usepackage{algorithm}
\usepackage{algpseudocode}

\onlineid{0}

\newcommand{\system}{SIA}

\vgtccategory{Research}

\title{%
    Linking Heterogeneous Data with Coordinated Agent Flows\\
     for Social Media Analysis
}

\author{%
  Shifu~Chen,
  Dazhen~Deng,
  Zhihong~Xu,
  Sijia~Xu,
  Linyu~Qin,
  Tai-Quan~Peng, and
  Yingcai~Wu
}

\authorfooter{

  \item
    Shifu Chen, Dazhen Deng, Zhihong Xu, and Sijia Xu are with the School of Software Technology, Zhejiang University.
    E-mail: shifu.chen@outlook.com, dengdazhen@zju.edu.cn, zhihong0114@gmail.com, xusijia2002@zju.edu.cn.

  \item
    Linyu Qin is with the College of Media and International Culture, Zhejiang University.
    E-mail: linyu.qin@zju.edu.cn.

  \item
    Tai-Quan Peng is with the Department of Communication, Michigan State University.
    E-mail: pengtaiq@msu.edu.

  \item
    Yingcai Wu is with the State Key Lab of CAD\&CG, Zhejiang University.
    E-mail: ycwu@zju.edu.cn.
}

\abstract{%
  Social media platforms generate volumes of heterogeneous data, capturing user behaviors, textual content, and network structures. Analyzing such data is crucial for understanding phenomena such as opinion dynamics, community formation, and information diffusion. However, discovering insights from this complex landscape is exploratory, conceptually challenging, and requires expertise in social media mining and visualization. Existing automated approaches, including large language models (LLMs), remain largely confined to structured tabular data and cannot adequately address the heterogeneity of social media analysis.
We present SIA (Social Insight Agents), an LLM agent system that links heterogeneous multi-modal data, including raw inputs (e.g., text, network, and behavioral data), mined analytical results, and rendered visual artifacts, through coordinated agent flows. Guided by an insight-oriented taxonomy connecting insight types with suitable mining methods and visualization strategies, SIA adopts a stage-synchronized strategy that proceeds through goal decomposition, query, mining, visualization, and reporting stages. At each stage, it collects prior information to jointly plan and execute agent actions, while the coordinator maintains cross-stage action dependencies and assembles and distributes data to agents. Through quantitative evaluation and case studies supported by an interactive interface, we show that SIA can discover diverse and meaningful insights from social media with opportunities for subsequent reliability assessment.
}

\begin{document}

\maketitle
\section{Introduction}
Social media platforms generate volumes of data every day, capturing diverse opinions, rich interactions, and dynamic trends. Effectively analyzing such complex and massive data is crucial for understanding public opinion dynamics~\cite{OpinionFlow, Whisper}, information diffusion~\cite{DMap, RMap}, and topic evolution~\cite{TopicCompetition, EvoRiver}. 
Visual analytics, which integrates advanced data mining algorithms~\cite{DataminingSurvey} with intuitive visual representations~\cite{survey17}, has become indispensable for exploring these multifaceted social phenomena. 

However, visual analysis of social media data remains challenging due to its heterogeneity, scale, and evolving context. Analysts often need to integrate textual content, metadata, and interaction networks, apply suitable mining methods such as sentiment analysis, graph mining, and temporal pattern detection, and design visual encodings that communicate findings effectively. These steps are interdependent and typically require substantial domain expertise.

Recent advances in large language models (LLMs) have introduced a new paradigm for automatic visual analysis. Existing systems can decompose goals, execute mining operations, generate visualizations, and summarize insights. For example, InsightPilot~\cite{InsightPilot} and InkSight~\cite{InkSight} support insight discovery from tabular data, LightVA~\cite{LightVA} highlights agent planning in analytical workflows, and Data-Copilot~\cite{Data-Copilot} automates querying, processing, and visualization for large-scale datasets. Despite this progress, existing LLM-based systems still fall short of supporting general-purpose social media analysis.
\textbf{First}, aligning diverse analytical goals with fruitful methods in dynamic social media context is difficult: as social media tasks are diverse, ambiguous, and often evolve during exploration, it must translate high-level intents into suitable mining methods, data transformations, and visual encodings, which requires contextual knowledge about both the task and the analysis techniques.
\textbf{Second}, coherent integration of heterogeneous data sources within agent flows is challenging: 
text, metadata, temporal records, engagement signals, and interaction networks are complementary, but existing agent workflows often process them separately or assume pre-processed inputs, making it difficult to construct coherent analytical pipelines.
\textbf{Third}, supporting inspection and adjustment of multi-task, interdependent agentic workflows is challenging:
it must maintain cross-stage dependencies, as later outputs often rely on earlier artifacts such as queried data, mined patterns, and visualization specifications. However, these dependencies are often dynamic and implicit, making it difficult to reproduce workflows, restart from intermediate steps, or update affected downstream steps when an artifact changes.

Building on these challenges, we propose \textbf{Social Insight Agents (\system{})}, an LLM agent system for social media analysis. To support goal--method alignment, we introduces a social media insight taxonomy as an intermediate reasoning layer, helping the agent map high-level analytical goals to suitable mining methods and visualization strategies. To enable heterogeneous data integration, \system{} manages intermediate artifacts in a lineage graph, which serves as a shared data pool for tracking data dependencies and provenance. Built on this graph, a coordinator retrieves and recomposes relevant artifacts across query, mining, and visualization stages. To further evaluate the soundness and reliability of the workflow beyond final outputs, we use an interactive interface that exposes intermediate states and supports expert intervention. Finally, we evaluate \system{} through two real-world case studies and quantitative experiments.
In summary, our main contributions are:
\begin{itemize}[leftmargin=15pt, label=\(\diamond\)]
\item A \textbf{taxonomy of social media insights} that supports goal--method alignment in agent-driven social media analysis.
\item A \textbf{coordinated agent framework} that coordinates agents for query, mining, and visualization, together with lineage-based data management and coordination across stages.
\item A \textbf{comprehensive evaluation} through real-world case studies and quantitative experiments, supported by an \textbf{interactive interface} for inspecting and intervening analysis process.
\end{itemize}

\section{Related Work}
We review prior work in social media data mining, visual analytics for social media, and automatic insight discovery with LLM agents.

\subsection{Social Media Data Mining}

Social media data mining has developed diverse techniques for different data modalities and analytical targets. Text-oriented methods, such as topic modeling~\cite{TopicModeling}, sentiment analysis~\cite{Opinionmining}, and semantic structure analysis~\cite{SentenTree}, reveal public concerns, opinions, and discourse patterns. Network mining analyze user relations, including community detection~\cite{CommunityDetectionSurvey}, influence propagation~\cite{InfluenceMaximizationSurvey}, and bot detection~\cite{CACL}. Temporal mining detects trends or bursts, and behavioral changes over time~\cite{SES}, while multimodal analysis combines text, images, and user profiles to capture richer social context~\cite{Multiple}.
These methods provide essential building blocks for social media analysis. However, they are often tailored to specific data types or tasks, and choosing suitable methods still requires substantial domain knowledge. This challenges automated agent systems, which must connect high-level analytical goals with appropriate mining operations across heterogeneous data. Our work addresses this gap with a taxonomy of social media insight types that guides agents in selecting and organizing mining strategies.

\subsection{Visual Analytics for Social Media}

Visual analytics has been widely used in diverse analytical scenarios to explore and interpret social media data by combining computational analysis with interactive visualization~\cite{DBLP:conf/icwsm/PennacchiottiP11}. Network-oriented systems~\cite{DMap,Vizster} use node-link diagrams and community visualizations to reveal user relationships and group structures. Spatial-temporal systems~\cite{Whisper,TopicCompetition,EvoRiver} employ maps, flows, and timelines to analyze geographic distributions, information diffusion, and topic evolution. Text-oriented systems~\cite{OpinionFlow,RMap,survey17} use word clouds, semantic layouts, and topic-based visual encodings to help users understand discourse content and thematic structures. These systems demonstrate the effectiveness of visual analytics for making social media patterns interpretable. Nevertheless, most systems are designed for particular tasks or datasets with tightly coupled pipelines, which provide strong support for targeted analyses but are difficult to generalize to open-ended social media questions. In contrast, our work aims to support flexible visual analysis pipeline through an agentic workflow.

\subsection{Automatic Insight Discovery}

Automatic insight discovery aims to reduce the effort required to identify meaningful data patterns. Early systems mainly focused on predefined statistical insights, such as trends, correlations, outliers, and distributions. For example, QuickInsights~\cite{QuickInsights} and MetaInsight~\cite{MetaInsight} showed that automated methods can surface useful statistical regularities, but their predefined pattern spaces limit their ability to capture semantic context and diverse analytical intent.

Recent work has explored more flexible forms of insight discovery by incorporating natural language and LLMs. Systems such as InsightPilot~\cite{InsightPilot} and InkSight~\cite{InkSight} use LLMs to interpret data, generate explanations, and organize discovered insights. Chat2VIS~\cite{Chat2VIS} and LLM4Vis~\cite{LLM4Vis} further connect natural language queries with visualization, enabling users to express analytical intent more directly. Data-Copilot~\cite{Data-Copilot} and JarviX~\cite{JarviX} extend this direction by translating high-level analytical goals into executable data operations and visual outputs. More recently, agent-based systems such as LightVA~\cite{LightVA} have shown that multi-agent collaboration can support more adaptive insight discovery workflows.
Despite these advances, automatic insight discovery remains limited in social media analysis. Existing systems are primarily designed for structured datasets, where insights often correspond to statistical patterns over well-defined tables. In contrast, social media insights may involve content semantics, user behavior, temporal evolution, network structures, and cross-dimensional relationships among them. Discovering such insights requires agents to select appropriate mining methods, compose evidence from heterogeneous data sources, and present results in contextually meaningful visual forms. To address these requirements, our work introduces a taxonomy-guided agent framework for social media insight discovery, enabling agents to reason about insight types, coordinate multi-stage analysis, and manage heterogeneous data dependencies.

\begin{table*}[!ht]
\caption{Taxonomy of Social Media Insights}
\label{tab:design-space}
\centering
\small
\renewcommand{\arraystretch}{1.5}
\begin{tabularx}{\textwidth}{@{}c c >{\hsize=0.7\hsize}l >{\hsize=1\hsize\raggedright\arraybackslash}X >{\hsize=1\hsize\raggedright\arraybackslash}X@{}}
\toprule
\multicolumn{2}{c}{\textbf{Entity Types}} &
  \textbf{Insight Types} &
  \textbf{Static Insights} &
  \textbf{Dynamic Insights} \\
\midrule
\multirow{8}{*}{\textbf{User}}
  & \multirow{3}{*}{\textbf{Single User}}
    & Real-world Identity Attributes & Social, cultural, and political identity & Evolution of social, cultural, and political identity \\ 
  &  & Behavioral Signatures & User habits in content creation and interaction & Changes in user content creation and interaction habits \\ 
  &  & Digital Identity Attributes & Identity characteristics on social media platforms & Evolution of account identity characteristics \\
\cmidrule(l){2-5}
  & \multirow{5}{*}{\textbf{User Group}}     
    & Network Topology & User connections and their relative importance & Evolution of user relationships and social networks \\ 
  &  & Group Behavior Pattern & Interaction and activity patterns among users & Group-level behavioral patterns over time \\
  &  & Community Formation & Community structures and clusters of connected users & Emergence and evolution of user communities \\
      &  & Information Diffusion & Common routes of diffusion across network & Evolution of diffusion speed and coverage of network\\
  &  & Influence Center & Key users and groups with strong network influence & Influence propagation and cascade dynamics \\
\midrule
\multirow{6}{*}{\textbf{UGC}}
  & \multirow{3}{*}{\textbf{Single UGC}}
    & Content Features & Semantic and stylistic features of content & N/A: Content of a single post remains static \\
  &  & Contextual Metadata & Time, location, platform and device of content creation & N/A: Metadata of a single post remains static \\
  &  & Engagement Metrics & User interaction and responses to content & Shifts in content interaction patterns \\
\cmidrule(l){2-5}
  & \multirow{3}{*}{\textbf{UGC Group}}
    & Content Structure & Common themes, topics, and sentiment in content & Temporal evolution of topics and sentiment \\
  &  & Diffusion Distribution & Content distribution of time, location, and platform & Changes in temporal distribution patterns \\
  &  & Engagement Structure & User interactions across different types of content & Sudden bursts and shifts in interaction behavior \\
\bottomrule
\end{tabularx}
\end{table*}

\section{Heterogeneous Data in Social Media Analysis}
Heterogeneity in social media analysis arises from the raw inputs, the intermediate processing steps and the final visualization outputs.

\textbf{Input data.} Social media platforms generate massive, heterogeneous data that capture different aspects of user activity and interaction. Typical inputs include \textbf{tabular data} for structured attributes and quantitative measures (e.g., user profiles, demographics, and content metadata), \textbf{textual data} for semantic content and linguistic patterns (e.g., posts, comments, and replies), \textbf{visual data} for visual communication patterns (e.g., images and videos), and \textbf{network data} for relational structures and diffusion pathways (e.g., follower--followee relationships and user--content interactions). These inputs are linked by shared identifiers (e.g., user IDs and content IDs) and timestamps, enabling cross-source integration and coordinated analysis across modalities.

\textbf{Intermediate data.} Data mining and transforming further introduce heterogeneous intermediate results. Text mining may produce embeddings, semantic clusters, sentiment distributions, or syntactic structures; vision analysis may extract object recognition results, visual similarity matrices, aesthetic scores, or embeddings; network analysis may generate centrality scores, community structures, or diffusion cascades; and temporal mining may identify evolving trends, periodic cycles, or burst patterns. These results vary in format, granularity, and semantics, from numerical vectors and probability distributions to hierarchical clusters and graph representations, creating complex dependencies that must be coordinated to maintain analytical coherence across process pipelines.

\textbf{Visualization data.} Finally, the outputs of visual analytics---such as node-link diagrams, timelines, or topic maps---constitute another form of heterogeneous data, as visualization transforms analysis results into diverse visual elements and attributes. Through aggregation and encoding, data are mapped to node positions, edge weights, temporal bins, color mappings, scales, layout hierarchies, and other visual properties.

Handling such cross-stage heterogeneity requires mechanisms to consistently integrate and manage diverse data throughout the workflow. This motivates our \textbf{taxonomy} and \textbf{coordinated agent framework}, which guide unified heterogeneous data handling.

\section{Preliminary Study} \label{sect:Preliminary}
While LLM agents show promise in handling complex analytical tasks, applying them directly to social media insight discovery remains challenging without systematic guidance. To provide such guidance, we conducted a preliminary study involving two domain experts: E1 holds a Ph.D. in communication, with research interests in computational social science and political communication; and E2 specializes in cybersecurity and conducts research on online public opinion analysis, bringing expertise in interpreting patterns of information spread and manipulation. Through multiple rounds of discussion, the experts helped refine the taxonomy and requirement analysis, ensuring both theoretical rigor and practical applicability. The study protocol was reviewed and approved by our lab’s internal research review process.

\subsection{Social Media Insight Taxonomy}
A key challenge in supporting LLM-based social media analysis is to provide explicit guidance on what kinds of insights to seek.

\textbf{Need for an Insight-Oriented Taxonomy.}
Prior taxonomies~\cite{DataminingSurvey, Survey16, survey17, bigdata, techniquessurvey} provide valuable bottom-up organizations of the social media analysis and visualization design space, covering data types, analytical operations, visualization tasks, chart types, and mining methods.
These perspectives are useful for comparing data types, analytical methods, and visualization forms, but technique-centered categories do not always reflect how users formulate questions in natural-language interactions with LLMs. Although users may explicitly specify low-level analytical operations, such as ``run community detection, sentiment classification, and temporal aggregation, then visualize the results using a node-link diagram and a timeline,'' they often express the same analytical need more naturally as a higher-level question: ``Who are the main groups discussing this event, what positions do they take, and how did the discussion evolve over time?''
Brehmer et al.'s typology~\cite{6634168} is closer by describing abstract actions such as discovering, comparing, and summarizing, yet these labels still do not indicate which mining methods should follow: ``discover'' may refer to communities, topics, stance patterns, diffusion pathways, or temporal shifts.
Insight types provide the missing link: they are categories of human-readable descriptions of analytical results, such as group structure, stance distribution, or temporal evolution.
We therefore decompose such questions into insight types and then map them to mining methods or visual encodings.
Thus, \system{} adopts an insight-oriented taxonomy as a semantic layer between natural-language goals and executable analysis specifications.

\textbf{Literature Collection.}  We first collected papers referenced in two comprehensive and influential survey studies on social media visual analytics~\cite{survey17, Survey16}. To expand our coverage of data mining methodologies, we additionally incorporated references from social media data mining surveys~\cite{DataminingSurvey} as supplementary sources. We then further expanded our corpus through additional papers from major visualization and data mining venues (IEEE TVCG, IEEE TKDE, JOV, IEEE VIS, EuroVis, PacificVis, KDD, WWW, WSDM) between 2010 and 2025. We prioritized works that presented detailed case studies or usage scenarios. Additional references were gathered using backward citation tracing and keyword searches in academic databases, applying the same extraction procedure. The final collection consists of 34 papers.

\textbf{Corpus Extraction.}
For each paper, we examined its case studies and extracted three elements: insight descriptions, mining techniques, and visualization examples. We represented insight descriptions and mining techniques as paired tuples, since they jointly define the analytical outcome. For example, a case study in \cite{TopicCompetition} identified how different actors attracted public attention to a social topic over time, which we paired with keyword extraction, temporal clustering, and trend analysis. Visualization examples were collected separately as an image corpus, as designs may vary even for similar analytical outcomes.

\textbf{Initial Taxonomy Construction.} Based on such extractions, we developed an initial taxonomy structure. We first organized insights by underlying entity type: \textit{users} and \textit{user-generated content (UGC)}, which are widely adopted across social media platforms. 
Second, we distinguished whether the analysis concerns individual entities or structural patterns among
multiple entities, resulting in four categories: single user, user group, single UGC, and UGC group. Third, we introduced a static--dynamic distinction to capture whether an insight describes stable characteristics or temporal evolution. Together, these dimensions ground our \(2 \times 2 \times 2\)  taxonomy.

\textbf{Iterative Refinement.} We then feed and iteratively refined insight types within each dimension through the following process:
\begin{itemize}[leftmargin=*, itemsep=3pt, topsep=3pt, parsep=0pt, partopsep=0pt]
\item Within each analytical entity category, we identified recurring patterns in insight textual descriptions and grouped them into distinct insight types based on their analytical focus and objectives.
\item For insight types involving temporal characteristics, we further subdivided them into finer-grained categories that capture different aspects of temporal evolution (e.g., trend and periodicity).
\item We extracted and mapped the corresponding data mining techniques for each identified insight type.
\end{itemize}

Throughout this process, we continually refined and merged similar insight types to improve consistency and coverage. After multiple iterations, we finalized a taxonomy of insight types (\autoref{tab:design-space}).

\textbf{Taxonomy Validation.} We conducted structured interviews with two domain experts to validate the taxonomy and assess its usefulness in guiding analysis. The validation had two parts. First, experts reviewed the taxonomy and discussed whether it missed important social media analysis problems. Second, each expert provided two recent analytical projects. For each project, we compared four analysis plans: expert's original plan, expert's plan reformulated with taxonomy, an LLM-generated plan without taxonomy guidance, and an LLM-generated plan with taxonomy guidance.
The result showed that the taxonomy was both comprehensive and useful for analysis planning. Expert feedback led to refinements that improved sociological relevance and technical feasibility. When using the taxonomy, experts identified related insight types that were not included in their original plans, suggesting that the taxonomy can broaden analytical coverage. They also found that LLM-generated plans became more structured and reasonable when guided by the taxonomy. Both experts further confirmed that social media analysis commonly follows a mining-to-visualization workflow, supporting our subsequent system design.

\subsection{Insight-Driven Visualization}
\begin{figure*}[t]
  \centering
  \includegraphics[width=\linewidth]{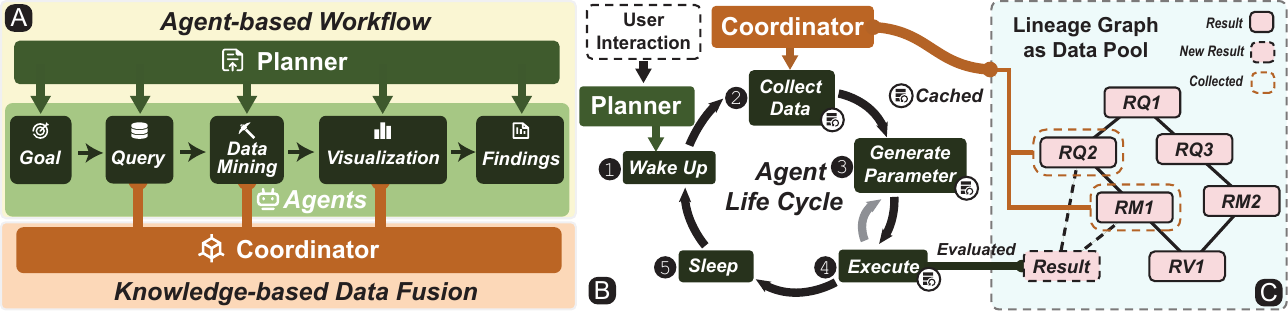}
    \caption{
    Overview of the agent-based workflow.
    Left: the planner coordinates multiple agents over heterogeneous social media data with the support of a heterogeneity coordinator.
    Right: analytical execution agents follows a lifecycle of wake-up, data collection, parameter generation, execution, and sleep, while post-execution user interactions can trigger agent re-execution through the planner.
    }

  \label{fig:overview}

\end{figure*}
To inform visualization design, we considered 1) experts' assessment of visualization types, including common charts and complex examples from the corpus, and 2) encoding channel choices from the corpus.

Experts made three observations. \textbf{First}, common charts such as bar, pie, and line charts remain important for communicating insights. \textbf{Second}, from complex corpus examples, they identified word clouds, graph visualizations, and stream graphs as useful for social media analysis, supporting textual content, network interactions, and temporal topic evolution. \textbf{Finally}, they emphasized multidimensional insight integration, as effective social media analysis often requires combining multiple dimensions. As E1 noted, ``the core of social media analysis lies in understanding the relationships between multiple dimensions.''

In parallel, we lightly annotated encoding channels in the visualization corpus. For graph-based views, color may distinguish communities or user groups, position reveals network structure, size indicates importance or influence, and edges represent relationships or diffusion paths. We also annotated cross-view links through shared objects such as users, posts, topics, and time ranges. These observations informed visualization forms, encodings, and cross-view coordination.

\subsection{Requirement Analysis of Automatic Exploration}
\label{subsect:RequirementAnalysis}

We conducted a structured seminar with domain experts to identify system design requirements for automated social media analysis workflows. The discussion focused on practical challenges in multi-source analysis and the potential role of LLM agents in supporting exploration.

\textit{C1: Multi-source Data Analysis.}
Based on the experts' need for multidimensional insight integration, we identified coordinated analysis across multiple social media data sources as a key challenge.

\textbf{R1: Data Subset Querying.}
Since social media analysis often targets specific subsets rather than entire datasets, the system must identify and retrieve relevant data across sources. It should support flexible query composition and efficient subset retrieval.

\textbf{R2: Data Heterogeneity Coordination.}
The system must coordinate heterogeneity across raw data, intermediate mining and transformation results, and visualization outputs, while preserving the relationships among these heterogeneous data forms throughout the process.

\textbf{R3: Multidimensional Insight Communication.}
Insights from coordinated analysis are often multidimensional and require visualization strategies that reveal cross-dimensional relationships. The system should map discovered insights to suitable visualization techniques and organize results into clear, communicable forms.

\textit{C2: Exploration Reliability.}
Although experts recognized the analytical potential of LLM agents, they expressed concerns about the reliability of automated exploration in practice, as automated insight discovery may produce inaccurate or misleading results.

\textbf{R4: Inspectable and Intervenable Discovery Process.}
To support expert oversight, the framework should maintain manageable execution states and clear relationships among exploration stages, enabling experts to inspect and intervene in the discovery process.

\textbf{R5: Mining Uncertainty Management.}
Experts highlighted that data mining processes inherently introduce uncertainty through algorithm choices, data selection, and parameter settings. The system should make such uncertainties explicit and examinable.

\section{Coordinated Agent Framework}
Our framework organizes analysis as an agent-based workflow while following basic visual analytics pipeline, which consists of three core components (\autoref{fig:overview}): \textbf{planner} serves as central controller that manages workflow execution and activates different analytical agents when needed; \textbf{analytical agents} are responsible for different analysis functions, including goal formulation, data retrieval (R1), analytical computation, visualization generation (R3), and findings synthesis; across these stages, \textbf{coordinator} manages data fusion, reuse, and dependency tracking across heterogeneous intermediate products (R2). Users can \textbf{interact} with the interface throughout workflow by reviewing and adjusting intermediate decisions (R4) and validating mining results (R5). Such interactions make intermediate results easier to inspect, while allowing the planner to trigger re-execution when necessary.

\subsection{Stage-synchronized Planner}

The planner advances analysis through five stages (\autoref{fig:overview}.A).

\textbf{First}, in goal analysis, it decomposes the user's analytical goal into concrete analysis directions.
\textbf{Second}, in data query, it retrieves relevant social media data subsets for each direction.
\textbf{Third}, in data mining, it applies suitable analytical operations to extract patterns from the retrieved data.
\textbf{Fourth}, in visualization, it presents the mined patterns through visual summaries.
\textbf{Finally}, in insight synthesis, it synthesizes the analytical and visual outputs into evidence-backed insights.
After task decomposition, we do not execute the resulting directions as fully independent workflows. These directions are often related: some query results can be reused across directions, and some mining results need to be visualized together to reveal their relationships. Therefore, the planner synchronizes at each stage, sharing accumulated stage contexts to coordinate transitions and reduce duplicate work. Thus, an analytical workflow transforms a high-level goal into structured insights through stage-synchronized planning and execution.

\subsubsection{Stage Planning}
For each stage, the planner constructs a stage plan (Algorithm~\ref{alg:stage-planning}) with four components: \(P\) for the topic-direction scope, \(O\) for required operations or organizational intents, \(C\) for coordination constraints, and \(E\) for expected outputs.
During \textbf{query} and \textbf{mining} stages, the planner first derives per-direction operations or methods and then merges and deduplicates them within each topic, while the coordinator assigns corresponding constraints. This is feasible because retrieval and mining tasks are relatively independent: each direction can often be specified separately and then integrated by removing redundant operations, such as retrieving tweets and user graphs, or applying topic modeling and community detection. \textbf{Visualization} planning differs because visualization objectives cannot be obtained through simple task decomposition. Since multiple mining results may describe complementary aspects of a phenomenon, the planner selects per-topic artifacts for visualization and relies on the coordinator to derive corresponding organizational intents and constraints. Thus, the visualization stage plan remains at a more abstract level, indicating which analytical dimensions or products should be communicated rather than prescribing fully specified tasks.

\subsubsection{Context Management}
To monitor execution progress, the planner maintains stage context consisting of current \textbf{stage plan} and accumulated \textbf{execution instances}. Formally, an execution instances represented as:

\begin{equation}
\textit{Instance}
=
\langle type,op,para,R,E,N\rangle.
\end{equation}
where \(type\) is the activated agent type, \(op\) the target operation, \(para\) the execution parameters, \(R\) the returned results, \(E\) the evaluation outcome, and \(N\) the suggested next actions.
The planner assesses stage coverage from accumulated execution instances. A stage is complete when all plan requirements are covered by accepted instances; otherwise, the planner continues activating agents for unfinished or failed operations.

\begin{algorithm}[h]
\caption{Stage Planning}
\label{alg:stage-planning}
\begin{algorithmic}
\Require User question $q$, stage $s$, existing artifact set $A$ 
\Require Topic-direction pairs $P=\{(T_i,D_i)\}_{i=1}^{n}$, $D_i=\{d_{i,j}\}_{j=1}^{m_i}$ 
\Require Query API set $\mathcal{Q}$, mining method set $\mathcal{M}$
\Ensure Stage plan $Plan=\langle P,O,C,E\rangle$

\If{$s$ is query} 
    \ForAll{$T_i$}
        \State $\{O_{i,j}\}_{j=1}^{m_i} \leftarrow \mathrm{Planner}(s,q,D_i,\mathcal{Q})$ \Comment{per-direction}
        \State $O_i \leftarrow \mathrm{Consolidate}(\{(d_{i,j},O_{i,j})\}_{j=1}^{m_i})$ \Comment{per-topic}
        \State $C_i \leftarrow \mathrm{Coordinator}(s,A,O_i)$
    \EndFor

\ElsIf{$s$ is mining}
    \ForAll{$T_i$}
        \State $\{M_{i,j}\}_{j=1}^{m_i} \leftarrow \mathrm{Planner}(s,q,D_i,\mathcal{M})$ \Comment{per-direction}
        \State $O_i \leftarrow \mathrm{Consolidate}(\{(d_{i,j},M_{i,j})\}_{j=1}^{m_i})$ \Comment{per-topic}
        \State $C_i \leftarrow \mathrm{Coordinator}(s,A,O_i)$
    \EndFor

\ElsIf{$s$ is visualization}
    \ForAll{$T_i$}
        \State $V_i \leftarrow \mathrm{Planner}(s,q,T_i,A)$ \Comment{per-topic vis. artifacts}
        \State $(O_i,C_i) \leftarrow \mathrm{Coordinator}(s,V_i)$
    \EndFor
\EndIf

\State $O \leftarrow \{O_i\}_{i=1}^{n},\ C \leftarrow \{C_i\}_{i=1}^{n}$
\State $E \leftarrow \mathrm{Planner}(s,P,O,C)$
\State $Plan \leftarrow \langle P,O,C,E\rangle$
\State \Return $Plan$
\end{algorithmic}
\end{algorithm}

\subsection{Core Analytical Agents}

To support agent-driven analysis of social media data, we define five core analytical agents corresponding to the stages: goal analysis, query, data mining, visualization, and findings overview.

\subsubsection{Interface Agents}
Interface agents mediate between the user and the analytical workflow, translating user-facing inputs and outputs into internal representations for the planner and analytical agents.
\textbf{Goal Agent} converts the user's analytical request into a structured representation for planning. Using our insight \textit{taxonomy}, it identifies main topics and analytical directions, where topics support query planning and directions guide query construction, mining selection, and visualization design.
\textbf{Report Agent} turns accepted analytical results and rendered visual outputs into a user-facing findings overview. It organizes findings according to the user's goal, links them to visual evidence, and summarizes key patterns, trends, and insights in an interpretable form.

\subsubsection{Analytical Execution Agents}
In contrast to interface agents, the query, data mining, and visualization agents execute the core data-dependent operations of the workflow.

\begin{figure*}[t]
  \centering
  \includegraphics[width=\linewidth]{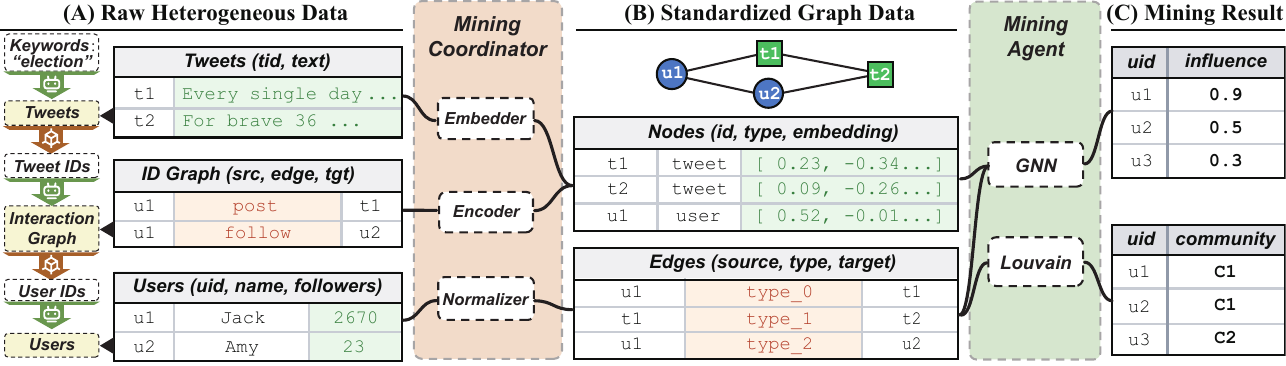}
    \caption{Query--mining coordination. Query and mining coordinator bridges agents by converting upstream agent's output artifacts into forms required by downstream agents. 
(A) Within a query chain, agents and coordinator operate alternately: each query agent produces intermediate artifacts, and the coordinator reformats them as inputs for subsequent query agents.
(B) For mining, the mining coordinator further integrates heterogeneous query artifacts by embedding textual content, encoding interaction relations, and normalizing user attributes into standardized graph data with typed nodes, edges, and feature embeddings.
(C) Mining agents consume the standardized representation to run analyses such as influence estimation or community detection. Throughout this process, artifact dependencies and data flows are recorded in lineage graph.
}
  \label{fig:mining}
\end{figure*}

\textbf{Execution Model.}
These agents follow a unified data-dependent lifecycle within a stage-wise workflow (\autoref{fig:overview}.B) and can be activated multiple times as execution instances within each stage. Once woken up by the planner, an agent \textit{collects data} from the coordinator-managed data pool, \textit{generates execution parameters} based on the current task and available upstream results, and \textit{executes} the corresponding operation. During execution, the agent evaluates whether the result satisfies task requirements. If fails, the agent returns to parameter generation and re-executes the operation with revised parameters. Otherwise, the agent returns the result to the data pool (\autoref{fig:overview}.C) and enters \textit{sleep mode} retaining input, parameters, output,  and evaluation results in \textit{cache}, where it remains available for later inspection and can be reactivated when new tasks or user interventions arise.

\textbf{Query Agent} is responsible for retrieving relevant data subsets from heterogeneous sources to support the current analytical task:
\begin{equation}
\text{subset} = f_{\textit{t}, \textit{cfg}}(source)
\end{equation}
where $\mathcal{D}$ denotes the available social media data source, type \textit{t} is determined by the query type, and configuration \textit{cfg} specifies retrieval constraints such as source, time range, keywords, or entity filters. The resulting subset is returned to the coordinator and used as input for downstream query, data mining and visualization agents.

\textbf{Data Mining Agent} applies analytical algorithms to data subsets to uncover patterns, communities, and predictive structures.
\begin{equation}
\text{pattern} = f_{\textit{algo}, \textit{cfg}}(\text{subset})
\end{equation}
where algorithm \textit{algo} is selected with taxonomy, and configuration \textit{cfg} is set by default.
Patterns are evaluated through a weighted combination of domain-specific metrics and LLM-based assessment.
\begin{equation}
E_m = \lambda_1 S_{\text{stab}} + \lambda_2 S_{\text{metric}} + \lambda_3 S_{\text{LLM}}.
\end{equation}
where \(S_{\text{stab}}\), \(S_{\text{metric}}\), \(S_{\text{LLM}}\) measure run-to-run stability, algorithm-specific quantitative quality (e.g., modularity for community detection or accuracy for prediction), and LLM-assessed interpretability, respectively. The weights \(\lambda_1\), \(\lambda_2\), and \(\lambda_3\) are algorithm-specific and configurable, reflecting the relative importance and availability of these scores. We set default weights empirically from pilot runs and manual inspection to align the aggregated score with human quality judgments. Uncertainty is then quantified by combined suitability and reliability:
\begin{equation}
U_m = \lambda_4 U_{\text{algo}} + \lambda_5 (1 - E_m).
\end{equation}
where \(U_{\text{algo}}\) denotes algorithmic uncertainty score, capturing uncertainty introduced by the mining method itself, such as stochastic outputs or limited reliability of available metrics.
\(\lambda_4\) and \(\lambda_5\) balance algorithmic and evaluation uncertainty, respectively.
In our implementation, we empirically set \(\lambda_4 = \lambda_5 = 0.5\), treating two sources equally.
These defaults are fixed during reported experiments.

\textbf{Visualization Agent} transforms mining results into executable visualization programs that communicate patterns, structures, and trends to users. It realizes visualization through code generation:
\begin{equation}
\text{code} = f_{\textit{t}, \textit{cfg}}(\text{pattern}), \qquad \text{view} = g_{\mathrm{render}}(\text{code})
\end{equation}
where type \textit{t} and configuration \textit{cfg} are determined jointly by taxonomy guidance and coordination. The taxonomy narrows the design space by matching insight structures such as temporal evolution, ranking, and community structure to candidate view families, while the coordinator determines whether multiple mining results should be communicated through an integrated view or through coordinated views. The generated code is executed by a deterministic renderer to obtain the final view. The quality of a view is assessed by LLMs following LIDA~\cite{dibia-2023-lida}.

\subsection{Coordinator}
We introduce a lineage-driven coordinator to manage non-linear data flow among agents and preserve traceable dependencies across stages.

\subsubsection{Agent Data Flow}
Although our workflow is organized as a pipeline, the data flow among agent actions is networked rather than strictly linear. Agent actions do not simply pass outputs to the next action in sequence. Instead, each action may draw on multiple outputs produced by previous actions as inputs, constraints, or context, while a single output may be reused by multiple subsequent actions. These many-to-many dependencies allow outputs to be branched, reused, and recombined across the workflow.

To coordinate this networked flow, the coordinator maintains a lineage graph as its storage structure. Nodes represent outputs produced by agent actions, and edges capture their production and dependency relations. This graph enables the coordinator to locate reusable outputs, assemble inputs for downstream agents, avoid redundant computation, and trace final outcomes back to their sources.

\subsubsection{Query and Mining Coordination}

Query and mining coordination bridges agents by managing how downstream inputs are extracted from upstream data products and how newly generated products are added to data pool (\autoref{fig:overview}.C).

\textbf{Extraction.}
For query operations, coordinator extracts fields such as tweet or user identifiers as inputs for subsequent query agent; for example, tweets retrieved by a topic keyword provide tweet identifiers for interaction retrieval, and the resulting interaction graph can further provide user identifiers for user retrieval, as shown in \autoref{fig:mining}.A.
For mining operations, it reshapes heterogeneous query products into the input forms required by downstream mining methods.
For example, textual content may be passed to an embedder, numerical attributes may be normalized, and categorical fields may be encoded into numerical features, so that the resulting representations can be used by methods such as community detection or GNNs, as shown in \autoref{fig:mining}.B.

\textbf{Lineage Construction.}
After query or mining agents generate new products, the coordinator inserts them into the lineage graph and establishes dependency relations according to the output structure of the corresponding operation.
When shared identifiers are preserved, such relations can be directly constructed through identifier correspondence, as in interaction graphs that retain tweet and user identifiers.
When mining outputs introduce higher-level abstractions, the coordinator constructs additional mappings based on the method-specific semantics of the output to preserve traceability.
For example, community assignments can be linked to users through node identities, while topics discovered from text collections can be associated with their supporting posts.
In this way, both low-level retrieval products and high-level analytical products remain connected to their evidence sources.

\begin{figure*}[tb]
\vspace{-5em}
  \centering
  \includegraphics[width=\linewidth]{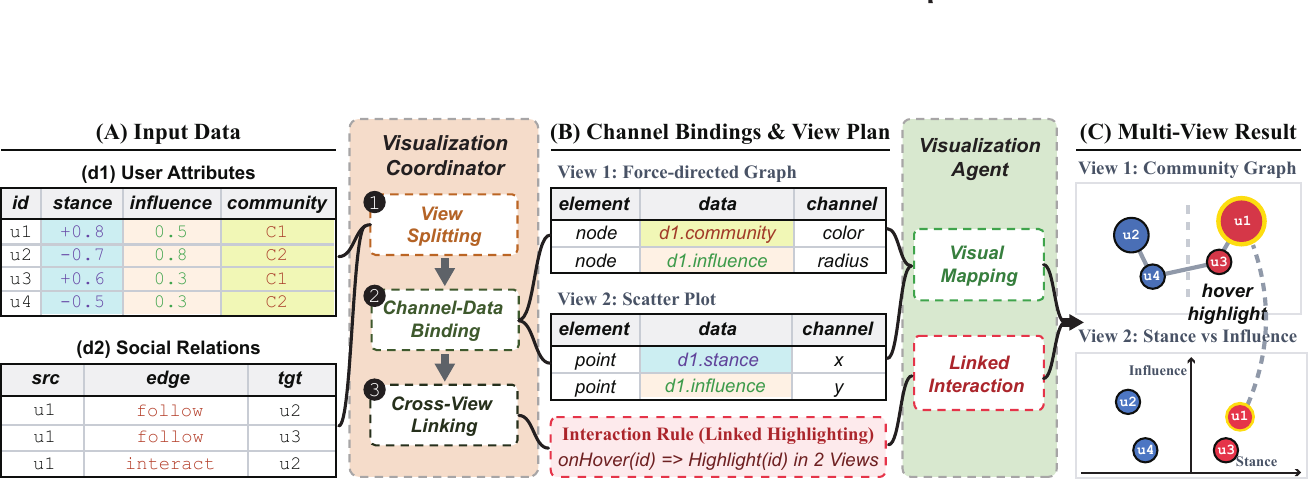}
  \caption{Visualization coordination. (A) Analytical products retrieved from data pool, including user attributes and social relations. (B) The coordinator organizes these products into visualization-ready structures through channel binding, view composition, and view splitting. Products that can be jointly represented are merged into a single view through visual mappings, while incompatible products are assigned to separate views. Dependency relations are further transformed into cross-view interaction specifications such as linked highlighting. (C) The resulting coordinated multi-view visualization preserves analytical coverage while maintaining traceable connections to underlying analytical products.}
  \label{fig:vis}

\end{figure*}

\subsubsection{Knowledge-Based Visualization Coordination}

Multiple mining results often need to be presented together because they describe related patterns from different analytical perspectives. The planner specifies a high-level visualization goal rather than concrete view organization, encodings, or interactions. The coordinator then turns this goal into a multi-view specification by deciding which products define view structures, which enrich existing views, and which lineage relations connect views and preserve traceability.

\textbf{Design Principles.}
Coordination follows three principles when refining the planner's goal.
\textit{Coverage} includes complementary mining products and preserves necessary correspondences.
\textit{Economy} avoids redundant views or encodings by merging products with shared visual structures.
\textit{Compatibility} combines products only when their analytical units, data structures, and encodings can be interpreted coherently.

\textbf{View Organization and Channel Binding.}
The coordinator organizes views and assigns visual channels in three steps:

\textit{(a) Initialize group-level views.}
It first identifies group-level mining products, such as user communities, interaction networks, or keyword groups, because they often require more structured visualization designs and are well suited for providing an overview of the analysis results. It then maps each product to a suitable visualization type based on its semantics and data form, e.g., force-directed graphs for user interactions (\autoref{fig:vis}.B) and word clouds for UGC keyword patterns. These group-level products usually occupy the primary structural channels of a view, such as nodes and links in a network, time windows in a timeline, or rows and columns in a matrix. After initializing the view structure, it records the remaining visual channels that can be used for further enrichment, such as color, size, position, labels, or annotations.

\textit{(b) Enrichment with single-level products.}
After initializing group-level views, the coordinator traverses the lineage graph to identify related single-level mining products (\autoref{tab:design-space}), such as user attributes, influence scores, or sentiment labels. These products provide detailed information about individual users or UGCs and can enrich the overview. It then binds them to remaining non-structural visual channels only when their analytical units, value types, and encodings are compatible with the target view. For example, categorical attributes are mapped to color or shape; numerical scores to size, length, opacity, or ordering; and textual evidence to labels, annotations, or tooltips.

\textit{(c) Handle unmatched single-level products.}
For single-level products that cannot be embedded into existing group-level views, the coordinator attempts to form additional views. It first matches compatible products pairwise based on their analytical units and value types. For example, two numerical scores over the same users or UGCs can be combined into a scatterplot (\autoref{fig:vis}.B). Products that cannot be paired are then presented as compact aggregate views according to their structure, such as bar charts for ranked categories, pie charts for part-to-whole proportions, or summary tables for detailed evidence.

\textbf{Lineage-Based Cross-View Linking.}
Finally, it builds cross-view links through lineage graph. During visualization construction, each visual element is linked to its upstream mining results, which are further linked to query-stage entities such as tweets and users. These relations form provenance paths from visual elements back to the original data. The coordinator identifies cross-view correspondences by checking whether two visual elements share valid provenance under predefined linking rules. For example, two elements can be linked through hover highlighting if they refer to the same user (\autoref{fig:vis}.C).

\section{Evaluation}
\begin{figure*}[tb]
  \centering
  \includegraphics[width=\linewidth]{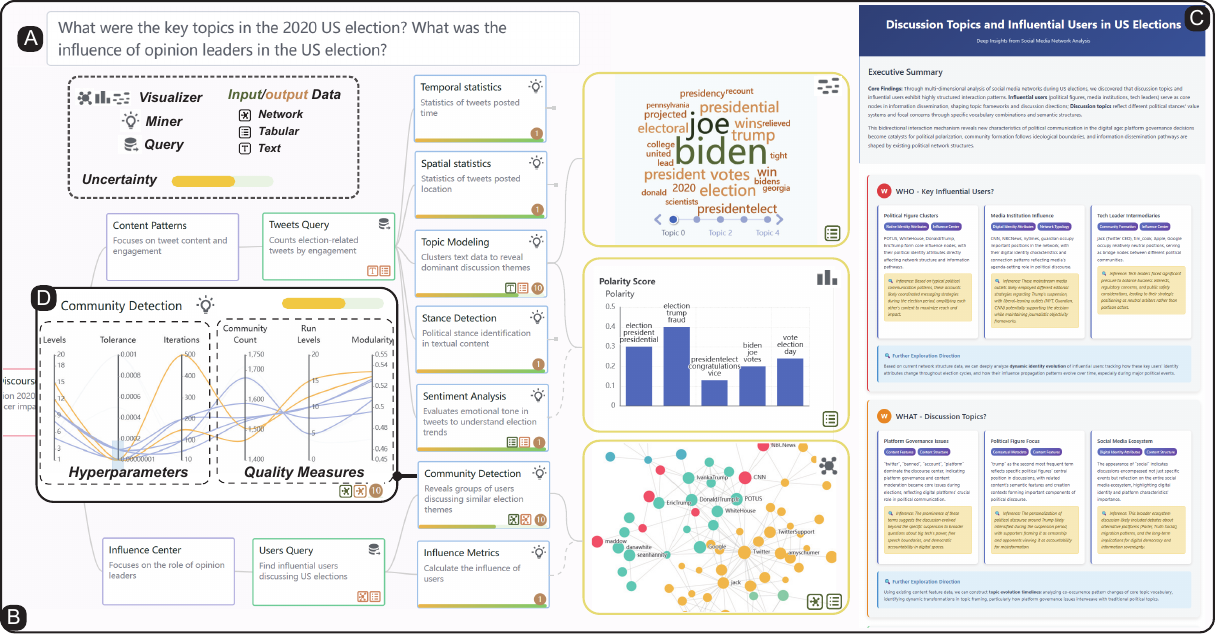}
  \caption{System interface. Chat Panel (A) facilitates dialogue between users and system. Action View (B) selectively displays the agent's actions during insight discovery. Findings Overview (C) organizes discovered findings in a taxonomy-guided documents and links them to their supporting views. Mining View (D) visualizes relationships between hyperparameters and corresponding results within a Miner node.
}
  \label{fig:case1}
\end{figure*}
We evaluate \system{} through expert-centered case studies, insight quality assessment, and LLM action performance measurement, which was approved through our lab's internal research review process.
\subsection{Global Setup}
\label{subsec:global}

We used a unified setup across all evaluation activities, covering system implementation, expert involvement, and task construction.

\textbf{Database.}
We uses TwiBot-22~\cite{Twibot22}, a large-scale Twitter dataset containing 1M users, 80M tweets, and a heterogeneous graph including user--user, tweet--tweet, and user--tweet relations.

\textbf{Expert involvement.}
We involved the same two domain experts from the preliminary study, using consistent identifiers throughout the paper. They provided requirements, domain background, representative questions, expected insights, and visualization needs, but did not participate in the technical design, including planning, coordination, agent design, evaluation, or interface implementation.

\textbf{Task construction.}
Before evaluation, each expert proposed 5--10 research questions within the time span of TwiBot-22~\cite{Twibot22}. After merging overlaps and checking data sufficiency, we retained 10 representative questions as the common task set.

\subsection{Quantitative Performance Evaluation}

We evaluate whether different LLMs can effectively support our agent framework, including planner and analytical execution agents

\textbf{Setup.}
We use five literature-derived analytical questions within the temporal scope of TwiBot22~\cite{Twibot22}, and run each question three times for each model.
We compare Qwen3-30B-A3B, GPT-5, DeepSeek-V3.1 without deep thinking mode, and GPT-4.1.
For each model, we sample 1,000 planner actions and 1,000 invoke (parameter generation) actions.

\begin{table}[h]
\centering
\caption{Comparison of response time and error rate across LLMs.}
\label{tab:qstudy}
\footnotesize
\renewcommand{\arraystretch}{1.28}
\setlength{\tabcolsep}{2.5pt}
\begin{tabularx}{\columnwidth}{@{}
>{\hsize=1.55\hsize\centering\arraybackslash}X
>{\hsize=0.90\hsize\centering\arraybackslash}X
>{\hsize=0.90\hsize\centering\arraybackslash}X
>{\hsize=0.825\hsize\centering\arraybackslash}X
>{\hsize=0.825\hsize\centering\arraybackslash}X
@{}}
\toprule
\multirow{2}{*}{\textbf{Model}} 
& \multicolumn{2}{c}{\textbf{Response Time (s)}} 
& \multicolumn{2}{c}{\textbf{Error Rate (\%)}} \\
\cmidrule(lr){2-3} \cmidrule(lr){4-5}
& \textit{Plan} & \textit{Invoke} & \textit{Plan} & \textit{Invoke} \\
\midrule
\textbf{Qwen3-30B-A3B}  & $9.8{\pm}3.5$  & $4.3{\pm}3.8$   & 2.2  & 7.9  \\
\textbf{DeepSeek-V3.1}  & $17.2{\pm}6.5$  & $5.8{\pm}5.0$    & 11.5 & 1.6  \\
\textbf{GPT-4.1}        & $\mathbf{4.2{\pm}1.3}$   & $\mathbf{1.7{\pm}0.7}$    & 6.0  & 11.5 \\
\textbf{GPT-5}          & $53.1{\pm}22.3$ & $21.4{\pm}14.2$  & \textbf{1.9}  & \textbf{0.0}  \\
\bottomrule
\end{tabularx}

\end{table}

\textbf{Metrics.}
\textit{Response time} indicates whether the prompt, context, and expected output length are practical for interactive use.
\textit{Error rate} measures the proportion of actions whose outputs fail to satisfy input schema or dependency constraints.
An action is counted as erroneous if it violates argument specifications or triggers an execution error due to missing or inconsistent prerequisites.

\textbf{Result.}
All four models keep error rates below 12\%, and most failed actions are corrected within two retries.
GPT-5 has the lowest error rate and the longest response time.
GPT-4.1 is the fastest while maintaining acceptable reliability, so we use it as default.

\subsection{Expert-Centered Case Studies}
We conducted case studies with both experts, each session lasted about one hour.
We pre-ran \system{} to generate initial results for the questions.
During the studies, the experts reviewed the outputs, interacted with the interface, and refined their interpretations.

\subsubsection{Interface}
The interface comprises four coordinated views: \textbf{Chat Panel} (\autoref{fig:case1}.A) is used for posing questions.
\textbf{Agent Tree} (\autoref{fig:case1}.B) presents the analysis workflow as a hierarchical structure. Each node represents an agent instance, and each edge denotes an execution dependency. The tree captures the stage-by-stage progression of the workflow: root nodes denote user goals, direction nodes represent decomposed analytical directions, and execution nodes correspond to query, mining, and visualization action.
\textbf{Mining View} (\autoref{fig:case1}.D) visualizes all explored mining attempts in a parallel coordinates plot, revealing how sensitive the mining result is to parameter changes and the stability of the mining process.
Hyperparameter dimensions and quality measures are aligned horizontally, and each line represents one parameter configuration.
User can open the \textit{Mining View} to either examine existing parameter sweeps or add new parameter settings for re-execution.
\textbf{Findings Overview} (\autoref{fig:case1}.C) provides a compact summary of discovered findings across multiple directions and offers an interaction hub that connects these findings back to their supporting visual evidence.

\subsubsection{Case I: American Election 2020}

In this case study, an expert aimed to explore key discussion topics during 2020 U.S. election and examine the influence of opinion leaders. \system{} first decomposed the goal into two directions, namely content patterns and influence centers, generated topic-specific keywords and then executed the analysis stage by stage.

At \textbf{query stage}, the planner built a plan to cover the data required by both directions and organized the retrieval process into multiple interdependent steps. At \textbf{mining stage}, the planner assigned mining methods according to two directions, including LDA-based topic modeling, sentiment analysis, stance detection, community detection, and influence metrics.
At \textbf{visualization stage}, the planner selected representative mining results for presentation, and the coordinator refined the view designs and visual encodings to accommodate these results. The topic modeling and sentiment analysis outputs were combined into a topic-sentiment bar chart, which compared the polarization degree across topics. Meanwhile, the community detection and user influence outputs were combined into an influence-aware community graph, where user influence was encoded by node size. These led to two insights. First, the word clouds showed that one topic centered on election fraud, the sentiment/topic bar chart further showed that this topic had stronger emotional polarization than others, suggesting Trump's allegations of Democratic vote fraud were associated with more divisive discussions. Second, the force-directed graph showed that many news media accounts were clustered with several Democratic figures and densely connected with them, suggesting that these media accounts were structurally closer to Democratic-related discussions and may have helped circulate narratives associated with that camp.

After reviewing the findings overview, the expert validated the community-level insights through the force-directed graph. By following the provenance links along the analysis chain, the expert traced the insight back to the mining view and added a preferred parameter setting with more iterations. The sleeping mining agent was then reactivated for re-execution. Since the resulting community structure and modularity changed little (yellow line), the expert regarded the result as stable and retained it for the analysis.

\subsubsection{Case II: COVID-19}

\begin{figure}[h]
  \centering
  \includegraphics[width=\columnwidth]{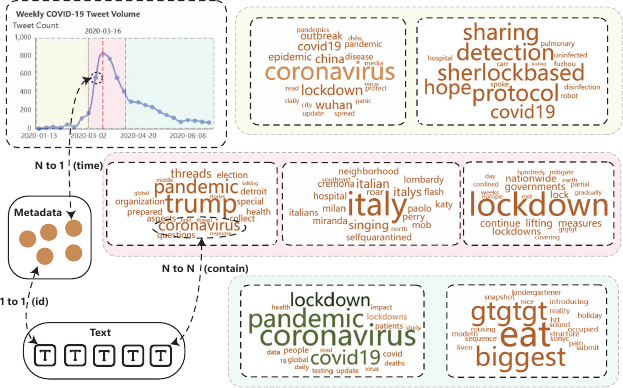}
  \caption{Temporal Analysis of COVID-19 social media discussions. The line chart shows weekly post volume of three distinct phases. Three wordcloud group revealing evolving topics across three phases. The coordination between these visualizations are established through linkages maintained in visualization coordinator.}
  \label{fig:case2}
\end{figure}

In this case study (\autoref{fig:case2}), the expert aimed to examine how COVID-19-related social media discussions evolved during the first half of 2020 (the latter half of 2020 contains sparse data in TwiBot-22). He started with the question: ``How did COVID-19 discussions on social media change over time in the first half of 2020?''
\system{} matched this inquiry to a dynamic content-pattern direction. 

At query stage, it retrieved COVID-19-related posts and performed preliminary temporal binning. At mining stage, change point detection identified a distinctive peak in mid-March, dividing the period into three phases: initial outbreak, peak phase, and declining phase. At visualization stage, the planner generated a weekly line-chart program to display these detected phases after rendering.

For these three phases, the mining stage executed topic modeling, and the visualization stage generated three sets of word-cloud programs that were rendered into views without duplicating shared visual specifications. Visualization Coordinator established multi-level data associations to enable cross-view coordinated interactions: first, words (visual elements) in the word clouds are connected to post texts through containment relation; post texts are then mapped to post metadata through unique IDs; finally, timestamps in the post metadata correspond to points (visual elements) in the line chart. This hierarchical data linking facilitates coordinated multi-view interactions: when users hover over words in the word clouds, it can trace back to posts containing those words and highlight the corresponding time points on the linechart; conversely, when users hover a time period on the linechart, the system can identify all posts within that timeframe, extract their key terms, and highlight them in the corresponding word clouds, thereby enabling bidirectional associative analysis between temporal patterns and content 
themes. Such coordinated views provide support for experts to conduct further exploration beyond reading the initial document.

\subsubsection{Expert Feedback}
\label{subsect:ExpertFeedback}

We collected expert feedback to reflect on experts' perceptions of the proposed automatic exploration workflow.

\textbf{Exploration Breadth and Coordination.}
Both experts commented that generated exploration directions covered many aspects they would consider in social media analysis. E2 noted, ``The agent thinks more comprehensively than we initially expected, covering exploration directions beyond my immediate consideration.'' The experts were particularly interested in the combinations of heterogeneous mining results. E1 commented, ``The agent appeared to explore an exhaustive range of combination methods. Despite some combinations appearing counterintuitive, it still revealed inspiring possibilities we hadn't considered.''

\textbf{Oversight and Trust in Automation.}
Experts also emphasized the need to keep the automated process understandable and controllable. They noted that automated mechanical insight-evaluation criteria alone are insufficient, because in sociological inquiry, the absence of salient insights can itself be informative, suggesting disorder, weak structure, or the lack of stable patterns. In particular, experts expressed concerns about using unfamiliar data mining methods without sufficient explanation. E2 commented, ``We usually employ simpler tools as alternatives, and unfamiliarity with the algorithms might become a barrier to trying new data mining methods.'' E1 further noted that they were interested in using language interaction to invoke complex data mining methods, especially for ``cross-validating them with our traditional methods.''

\subsection{Insight Quality Evaluation}
\label{subsec:insight_eval}

Following prior insight-based evaluation approaches~\cite{SaraiyaND05,North06}, we evaluate insight quality using expert ratings. We assess whether the visual outputs produced under different system conditions support useful and evidence-backed findings. Because open-ended social media analysis lacks a single exhaustive gold-standard insight set, we use a pooled expert-judgment protocol instead of predefined manual or LLM-generated ground truth. This avoids penalizing unexpected but defensible findings while leaving final validity to expert assessment.

\textbf{Setup.}
For each analytical question, evaluation proceeds as follows:
1) We execute the query stage once and freeze the retrieved dataset as a shared evidence base.
2) Full \system{}, \system{} without visualization coordination (w/o Coord.), and a Codex-based LLM-agent baseline independently generate visualizations from the same dataset.
3) To make heterogeneous visual outputs comparable, the same GPT-5.4 extractor is used as a condition-independent post-processing step to derive 2--4 candidate insight statements from each visualization, each consisting of a visual observation and an analytical implication as standard format.
4) All candidate insight statements are pooled, anonymized across conditions, and consolidated by merging statements that describe the same analytical conclusion into a single one.
5) Experts first rate the \textit{usefulness} of the anonymized statements according to their analytical value, and subsequently rate the \textit{confidence} of the insight units (statement with visual evidence) according to the extent to which the evidence supports the reported finding. Both dimensions are assessed using five-point Likert scales.
6) Ratings are aggregated across experts. For each analytical question, insight units are ranked by the mean of their aggregated usefulness and confidence scores. We report two threshold-based evaluation results, using the top 30\% and top 60\% ranked insight units as expert-endorsed insights, respectively.
 
\textbf{Metrics.}
We first report the mean \emph{usefulness} and \emph{confidence} scores across all candidate insights for each condition. We then compute \emph{precision} and \emph{recall} against the pooled expert-endorsed reference set. Specifically, \emph{precision} is the proportion of a condition's candidate insights that belong to the expert-endorsed set, indicating the quality of its generated insights. \emph{Recall} is the proportion of expert-endorsed insight units recovered by that condition, indicating its coverage of expert-endorsed findings.
Top 30\% threshold defines a stricter reference set of highly rated insights, while top 60\% threshold defines a broader set that also includes moderately endorsed but still useful insights. 

\begin{table}[h]
\caption{Insight metrics under different system conditions.}
\centering
\footnotesize
\renewcommand{\arraystretch}{1.16}
\setlength{\tabcolsep}{3.5pt}
\label{tab:insight_quality}

\begin{tabularx}{\columnwidth}{@{}
>{\centering\arraybackslash}p{0.19\columnwidth}
>{\centering\arraybackslash}p{0.17\columnwidth}
>{\centering\arraybackslash}p{0.17\columnwidth}
>{\centering\arraybackslash}p{0.17\columnwidth}
>{\centering\arraybackslash}p{0.17\columnwidth}
@{}}
\toprule
\multicolumn{2}{c}{\textbf{Metric}}
& \textbf{Full \system{}}
& \textbf{w/o Coord.}
& \textbf{Codex} \\
\midrule

\multicolumn{2}{c}{\# Visualizations}
& 95
& 135
& 100 \\

\multicolumn{2}{c}{\# Insights}
& 263
& 373
& 264 \\

\multicolumn{2}{c}{Avg. Usefulness}
  & \textbf{4.34 $\pm$ 0.47}
  & 3.68 $\pm$ 0.61
  & 3.37 $\pm$ 0.53 \\

\multicolumn{2}{c}{Avg. Confidence}
  & 4.08 $\pm$ 0.46
  & \textbf{4.44 $\pm$ 0.67}
  & 3.64 $\pm$ 0.68 \\

\midrule
\multirow{2}{*}{\textbf{\textit{Top 30\%}}}
& Prec.
& \textbf{42.21\%}
& 34.58\%
& 25.00\% \\

& Recall
& 39.64\%
& \textbf{46.07\%}
& 25.36\% \\
\midrule
\multirow{2}{*}{\textbf{\textit{Top 60\%}}}
& Prec.
& \textbf{93.54\%}
& 65.15\%
& 37.50\% \\

& Recall
& \textbf{43.46\%}
& 42.93\%
& 21.55\% \\
\bottomrule
\end{tabularx}
\end{table}

\textbf{Result.}
As shown in \autoref{tab:insight_quality}, full \system{} achieves the highest average usefulness and precision. Although it generates fewer visualizations and candidate insights than w/o Coord., its outputs are more frequently endorsed by experts, suggesting that visual coordination helps reduce redundancy and promote synthesized findings. w/o Coord. obtains the highest confidence score, which may reflect that its simpler, isolated encodings are easier for evaluators to inspect locally, even though they lead to lower precision.
For recall, w/o Coord. performs best at the top 30\%, while full \system{} slightly surpasses it at the top 60\%. This pattern suggests that full \system{} improves the quality of the broader ranked set, but does not always place all recoverable expert-endorsed insights at the very top.
The low insight overlap rate between full \system{} and Codex is only 14.42\%, indicating that our framework follows a substantially different exploration trajectory from pure LLM-based exploration. Together with Codex's lower precision and recall scores, these results suggest that directly prompting a general-purpose coding agent, without taxonomy-guided mining and visualization coordination, is less effective for this insight-generation task.

\section{Lessons Learned and Limitations}
This section reflects on the key lessons we gained during the development and evaluation of \system{}, and discusses its current limitations.

\textbf{Lineage as unified data management}.  
We found that lineage provides a natural abstraction for visual analytics workflows by viewing agent actions as transformations over analytical artifacts. Automated exploration rarely follows a linear sequence; instead, intermediate products are repeatedly reused and recombined to reveal different perspectives of the data, forming a many-to-many network of dependencies. Modeling these relationships explicitly unifies execution flow and data management within a single representation.

\textbf{Communication between LLM agent actions}.  
During development, we found that passing full historical context to LLMs often reduced focus. Our stage-synchronized communication mechanism, which summarizes results before stage planning, worked better for maintaining consistency and reducing redundant actions.

\textbf{Limitations}.  
First, our evaluation does not cover large-scale or real-time social media settings, and the prototype controls the amount of explored data and displayed visual elements. Future work could adopt mining and visualization modules tailored to large-scale and streaming scenarios.
Second, our implementation currently supports only hover highlighting for cross-view interaction, reflecting our focus on lightweight exploratory analysis. While the lineage graph could potentially support richer interactions such as filtering, selection propagation, or coordinated updates, these may trigger cascading effects that require more sophisticated propagation rules. Future work could explore such lineage-based interaction mechanisms.

\section{Conclusion}
We presented \system{}, a novel LLM agent system for analyzing heterogeneous social media data. By introducing a taxonomy of social media insights and a stage-synchronized coordinated framework that unifies planning and execution for querying, mining, and visualization, \system{} supports heterogeneous social media analysis across tabular, textual, and network data. Through expert-centered case studies and quantitative evaluation, we showed that \system{} can generate diverse and meaningful insights and support inspection and refinement of intermediate outputs in expert-facing workflows.

\acknowledgments{
The authors wish to thank A, B, and C. This work was supported in part by
a grant from XYZ.}

\bibliographystyle{abbrv-doi}

\bibliography{template}
\end{document}